\documentclass[conference]{IEEEtran}
\usepackage{amsmath,amssymb,amsfonts}
\usepackage{algorithm, algpseudocode, graphicx}

\newcommand{\x}{\mathbf{x}}
\newcommand{\y}{\mathbf{y}}
\newcommand{\z}{\mathbf{z}}

\newcommand{\B}{\mathbf{B}}
\newcommand{\F}{\mathbf{F}}
\newcommand{\bfH}{\mathbf{H}}
\newcommand{\bfP}{\mathbf{P}}
\newcommand{\G}{\mathbf{G}}
\newcommand{\R}{\mathbf{R}}

\newcommand{\bfeps}{\boldsymbol\epsilon}

\newcommand{\zero}{{\bf 0}}
\newcommand{\I}{{\mathbf I}}
\newcommand{\ones}{{\mathbf 1}}
\newcommand\numberthis{\addtocounter{equation}{1}\tag{\theequation}}

\title{Hierarchical Linear Dynamical System for Representing Notes from Recorded Audio}
\author{\IEEEauthorblockN{Leila Kalantari}
\IEEEauthorblockA{
\textit{University of Florida}\\
Gainesville, USA \\
leila@ufl.edu}
\and
\IEEEauthorblockN{Jose Principe}
\IEEEauthorblockA{
\textit{University of Florida}\\
Gainesville, USA \\
principe@cnel.ufl.edu}
\and
\IEEEauthorblockN{Kathryn E. Sieving}
\IEEEauthorblockA{
\textit{University of Florida}\\
Gainesville, USA\\
chucao@ufl.edu}
}
\begin{document}
%
\maketitle
\begin{abstract}
We seek to develop simultaneous segmentation and classification of notes from audio recordings in presence of outliers. The selected architecture for modeling time series is hierarchical linear dynamical system (HLDS).  We propose a novel method for its parameter setting. HLDS can potentially be employed in two ways: 1) simultaneous segmentation and clustering for exploring data, i.e.~finding unknown notes, 2) simultaneous segmentation and classification of audio recording for finding the notes of interest in the presence of outliers. We adapted HLDS for the second purpose since it is an easier task and still a challenging problem, e.g.~in the field of bioacoustics. Each test clip has the same notes (but different instances) as of the training clip and also contain outlier notes. At test, it is automatically decided to which class of interest a note belongs to if any. Two applications of this work are to the fields of bioacoustics for detection of animal sounds in audio field recordings and also to musicology.
 Experiments have been conducted for segmentation and classification of both avian and musical notes from recorded audio.
\end{abstract}

\section{Introduction}
\label{sec_intro}
High efficiency monitoring of biodiversity and ecosystem health is a priority in wildlife ecology. Many wildlife species (especially birds) are highly detectable by their sounds, therefore passive acoustic monitoring is appropriate to detect and survey such species. Yet, without accurate computer-automated retrieval of acoustic data, terabytes of recordings are useless. Our work can facilitate data retrieval strategies relying on signal processing and machine learning by helping eliminate the needs for manual recalibration, detector pre-training, and coping with variable acoustic conditions in the recording environment. 
\section{Technical background} \label{sec_bg}
\subsection{Kalman filtering}
Consider the following equations to model time series $\{\y_t\}$:
        \begin{align}
          \x_t &= \F_{t-1} \x_{t-1}+ \bfeps_t^\x \label{eq_hiddenLayer} & \mbox{ \ \ Hidden layer}\\
          \y_t &= \bfH_t \x_t     + \bfeps_t^\y \label{eq_obsLayer} & \mbox{ \ \ Observation layer}
\end{align}
Integer $t$ is the time index, vector $\x_t$ is the (hidden) state, vector $\y_t$ is the observation, $\F_t$ is the transition matrix, $\bfH_t$ is the observation matrix, $\bfeps^\x_t \sim \mathcal{N}(\mathbf{0}, \R^\x)$ is the uncertainty vector for the hidden layer and vector $\bfeps^\y_t \sim \mathcal{N}(\mathbf{0}, \R^\y)$ is the measurement error. 


Kalman filtering efficiently calculates $\hat{\x}_t$, an estimation of the current state $\x_t$ given the estimation of the previous state  $\hat{\x}_{t-1}$, previous information matrix $\bfP_{t-1|t-1}$ and the current observation $\y_t$ in an online fashion assuming the measurement and transition matrices are known for each time step.  Let $\I$ denote the identity matrix and
set the initial state and the initial information matrix as 
$\hat{\x}_0=(\bfH_0^T \bfH_0)^{-1} \bfH_0^T \y_0$ and 
$\bfP_{0|0}= (\bfH_0^T \bfH_0)^{-1}$ respectively.  One step of Kalman filtering is:
\begin{algorithm}[H]
\begin{algorithmic}[1]
\Function{KALMAN-step}{$\y_t, \hat{\x}_{t-1},\bfP_{t-1|t-1};\ \ \bfH_t, \F_{t-1}$}
   \State $\hat{\x}_{t|t-1} \gets \F_{t-1} \ \hat{\x}_{t-1}$\Comment{A priori state estimate}             
   \State $\bfP_{t|t-1} \gets \F_{t-1} \bfP_{t-1|t-1} \F_{t-1}^T + \R^\x$\Comment{A priori info matrix}
   \State $\G_t \gets \bfP_{t|t-1} \bfH_t^T (\bfH_t \bfP_{t|t-1} \bfH_t^T + \R^\y)^{-1}$\Comment{Kalman gain}     
   \State $\hat{\x}_t \gets \hat{\x}_{t|t-1} + \G_t (\y_t - \bfH_t \ \hat{\x}_{t|t-1})$\Comment{State update}
   \State $\bfP_{t|t} \gets (\I - \G_t \bfH_t) \bfP_{t|t-1}$\Comment{Info matrix update}
   \State \Return $\hat{\x}_t, \bfP_{t|t}$
\EndFunction
\end{algorithmic}
\end{algorithm}
\subsection{HLDS}
 HLDS models time series using a set of hierarchical linear systems  \cite{cinar2018hierarchical}.  Without loss of generality and for ease of presentation, assume $L=2$ where $L$ denotes the number of hidden layers.  The system of equations is:
\begin{align}
  \z_t &=  \z_{t-1}+ \bfeps^\z_t \numberthis\label{eq_org1} & \mbox{ \ \ 2nd hidden layer}\\
  \x_t &= \F \x_{t-1}+ \B \z_{t-1} + \bfeps^\x_t \label{eq_org2}& \mbox{ \ \ 1st hidden layer }\\
  \y_t &= \bfH \x_t + \bfeps^\y_t \numberthis\label{eq_org3}  & \mbox{ \ \ Observation layer}
  \end{align}
  where $\bfH$ is the observation matrix, $\F$ is the transition matrix, $\B$ is the coupling matrix, $\bfeps^\y_t \sim  \mathcal{N}(\zero,r^\y\I)$ is the measurement error,  $\bfeps^\z_t \sim \mathcal{N}(\zero,r^\z\I)$ is the uncertainty term for the second hidden layer, $\bfeps^\x_t \sim \mathcal{N}(\zero,r^\x\I)$ is the
 uncertainty term for the first hidden layer and $r^\y\I, r^\x\I, r^\z\I$ are covariances for the corresponding Gaussian distributions.  Assuming isotropic noise model for each layer is a simplification compared to the model described by equations (\ref{eq_hiddenLayer}) and (\ref{eq_obsLayer}) requiring less hyperparameters to be set (Section \ref{sec_noiseParams}).  
 
It is crucial that the transition matrix of the layer associated with $\z$ is identity. According to the model, time series $\{\z_t\}$ is unlikely to take a big leap over one time step if $r^\z$ is set small. This self-organizing property allows utilizing HLDS for the purpose of clustering.

Rewrite equations (\ref{eq_org1})-(\ref{eq_org3}) into canonical form as:
\begin{align}
\tilde{\x}_t &= \tilde{\F}\ \tilde{\x}_{t-1}+ \tilde{\bfeps}_t \label{eq_joint1} & \mbox{ \ \ Joint hidden layer}\\
          \y_t &= \tilde{\bfH}\ \tilde{\x}_t + \bfeps^\y_t \label{eq_joint2} & \mbox{ \ \ Joint observation layer}
\end{align}
where
  $\tilde{\x}_t =   \left(  \begin{matrix} \z_t \\ \x_t \end{matrix} \right)$ is the joint hidden state, 
  $\tilde{\F} = \left(  \begin{matrix} \I & \zero \\ \B & \F \end{matrix} \right)$ is the joint transition matrix,  $\tilde{\bfH} = \left(\zero \ \ \bfH\right)$  is the joint observation matrix and $\tilde{\bfeps}_t = \left(  \begin{matrix} \bfeps^\z_t\\ \bfeps^\x_t  \end{matrix} \right)$ is the joint uncertainty vector.    

\section{The proposed method}
\label{sec_meth}
We use the HLDS architecture from \cite{cinar2018hierarchical}.  HLDS imposes a constraint that effectively regularizes the Kalman specifically for note segmentation.  The time series data is modeled as a process that has multiple layers of hidden states. Higher layers have lower dimensions compared to lower ones. Also, higher levels exhibit longer term stationarity.  

In this paper, we propose to achieve the relative stationarity of layers in two novel ways: 1) Each layer is some low pass filtering of two consecutive states from the level just below it plus innovation (Section \ref{sec_paramSetting}).  2) 
Relative stationarity is also controlled by setting the innovation variances of higher layers smaller compared to lower ones (Section \ref{sec_noiseParams}). The specific block structure of the joint transition matrix pioneered by \cite{cinar2018hierarchical} provides efficient online updates (using Kalman filtering) to the hierarchy of hidden states of the process.

\subsection{Parameter setting}
\label{sec_paramSetting}
In HLDS, parameters are transition and coupling matrices for each hidden layer.  We propose to set them explicitly instead of training for them to avoid overfitting specially when training data is limited.  Parameters  of HLDS are set in a way that each hidden layer acts as low pass filtering of two consecutive (in time) states of the previous hidden layer (plus innovation modeled by Gaussian distribution).  Let $N$ and $S$ be the dimensions of $\x_t$ and $\z_t$ respectively and $N$ be divisible by $S$. The model is:
  \begin{align}
  \z_t &=  \z_{t-1}+ \epsilon^\z_t & \label{eq_prop1} \mbox{ \ \ 2nd hidden layer}\\
  \x_t &= - \x_{t-1}+ \B \z_{t-1} + \epsilon^\x_t \label{eq_layer1}& \mbox{ \ \ 1st hidden layer}\\
  \y_t &= \x_t + \epsilon^\y_t &\label{eq_prop3} \mbox{ \ \ Observation layer}
  \end{align}
  where
     $\B =  \left( \begin{matrix} \B_1, \dots, \B_S						 
    \end{matrix}\right)^T$. For each $s \in \{1, \dots, S\}$, block $\B_s$ is an $N/S \times S$ matrix whose $s$th column is $(2S/N)  \ones$ and zero elsewhere. The set of equations above can be rewritten in canonical form as in equations (\ref{eq_joint1})-(\ref{eq_joint2}) to be able to exploit Kalman filtering. 

\subsection{Hyperparameters setting}
\label{sec_noiseParams}
\subsubsection{Innovation and error parameters}
Other than our proposed parameter setting (Section \ref{sec_paramSetting}), the way we set innovation parameters contributes significantly to the relative stationarity of higher layer hidden states compared to lower ones.
There are $L$ Gaussian parameters to be set where $L$ is the number of the hidden layers.  They are denoted by $r^\z$ and $r^\x$ when $L=2$.  These hyperparameters should relate to each other in a way that none of the corresponding terms in $\lambda(\x_{t}) =(\y_t - \bfH_t \x_t)^T (\R^\y)^{-1}  (\y_t - \bfH_t \x_t) + (\x_t - \hat{\x}_{t|t-1})^T \bfP_{t|t-1}^{-1} (\x_t - \hat{\x}_{t|t-1})$  overwhelms another (the cost function \textproc{Kalman-step} implicitly solves for).  We achieved this by setting the innovation parameter of each layer proportional to the dimension of the layer.  We set $r^\y$, the error parameter, equal to be the innovation parameter of the bottom hidden layer in all our experiments.  We are aware that the best choice depends on the data.  However, we are able to achieve successful results in both of our experiments without considering the data (Sections \ref{sec_exp_avian} and \ref{sec_exp_trumpet}). 
\subsubsection{Number of hidden layers}
Another hyperparameter is $L$, the number of hidden layers.  In our experiment,  we set $L$ to $4$.   In future, we will study whether having more or less layers has any merits.
The minimum number of hidden layers necessary for the architecture of HLDS is two.
\subsubsection{Sliding time domain window length}
\label{sec_length}
It is desirable to decrease $w$, the length of the sliding window, to improve the computational performance.   However, $w$ should not be set too small. It has to be set large enough to contain enough information in a clip and small enough to maintain the stationarity assumption.  


\subsection{Training and testing}
\label{sec_mapping}
In the training step, observations $\{\y_t\}$ from each training note are mapped to the $\z$-space in the self-organizing manner implicated by HLDS model.  Note that ``training'' notes are only utilized for this step and parameters are set explicitly (Section \ref{sec_paramSetting}). 
These mappings are to be compared to where test observations get mapped to.
We still utilize Kalman filtering during training and testing to calculate states $\{\tilde{\x}_t\}$, associated with each train or test observation, in order to extract $\{\z_t\}$ as in \cite{cinar2018hierarchical}.  Mappings $\{\z_t\}$ can be exploited in many ways (next section).


\subsection{Post-processing}
\label{sec_post}
If $\z$-mappings from a training note form a single convex isotropic cluster, a simple distance-to-the-cluster-center heuristic in $\z$-space can be applied at the test time for the purpose of classification.  If the corresponding clusters have more complicated shapes as with avian notes, a secondary classification algorithm can be applied.  If the system is to be tested with outliers at the test time, a generative classification model can be applied in the $\z$-space \cite{coviello2010time, chan2009layered}. We applied kernel Mahalanobis distance for possibilistic classification \cite{PCKMD} for two reasons: It can discern outliers at the test time, and it has fully automated parameter setting.  Therefore, HLDS is the feature representation component in our prototypical classification system with PC-KMD as another component.   

\section{Experimenting with avian notes}
 \label{sec_exp_avian}

 The dataset of avian notes as used in our experiment is described here.
 Carolina chickadee ({\it Poecile carolinensis}) and tufted titmouse ({\it Baeolophus bicolor}) are the  two avian species are abbreviated by 
 {\tt pocaro} and {\tt babico} throughout the paper. Various notes uttered by {\tt pocaro} and {\tt babico} were recorded at $1$ m distance away from the bids.  The available labeled notes are
{\tt babico-D}, {\tt babico-F}, {\tt pocaro-Z}, {\tt babico-A},{\tt babico-Z}, {\tt pocaro-C}, {\tt pocaro-D} and {\tt pocaro-E}.  The number of instances of each note is $15, 6, 4, 4, 4, 6, 7$ and $10$ respectively for a total of $50$ instances.

\begin{figure*}[ht!]
\label{fig_birdExample}
\begin{tabular}{c}
\includegraphics[width=1.1\textwidth]{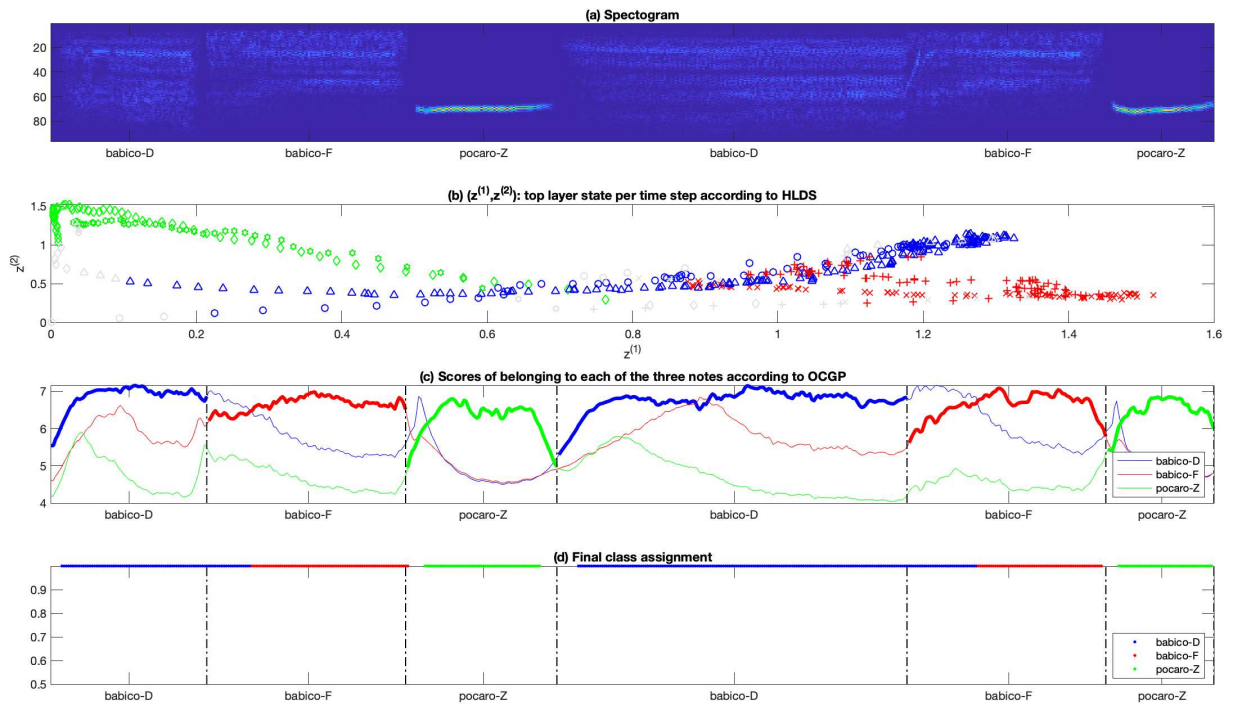}\\
\vspace{-1.25cm}
\end{tabular}
\caption{Our system on two instances of each of {\tt babico-D}, {\tt babico-F} and {\tt pocaro-Z} notes.  (a) The spectrum of the clip consisting of the six instances. (b) HLDS representation for each time step of the spectrogram. The transitionary states are blurred in gray to focus on inner states.  (c) Scores of each time step belonging to each of the three notes; the bolded scores are ideally above other scores in their corresponding segment (d) Final prediction for each time step.}
\end{figure*}

The training clip consists of two instances of each of the {\tt babico-D}, {\tt babico-F}, and {\tt pocaro-Z} (Figure \ref{fig_birdExample}(a)).
The test clip consists of the rest of the labeled data in a random order.   Matrix $\bfH$ is set to identity and the time domain sliding windows $\{\y^\prime_t\}$ are preprocessed as $\y_t = |dct( \y^\prime_t)|$ before applying HLDS.  Let $w^\prime$ and $w$ denote the dimensions of $\y^\prime_t$ and $\y_t$ respectively. The hyperparameters of HLDS are set as follows:  the dimension of hidden layers are $[96, 24, 12, 2]$ which implies $w^\prime=w=96$ since $\bfH$ is set to identity; $q=48$ where $q$ denotes the number of overlapping samples between two consecutive sliding windows.
Each of parameters $r^{\y}, r^{\x}, \dots,  r^{\z}$ is set proportional to the dimension of its corresponding layer as described in Section \ref{sec_noiseParams}.
We emphasize that all hyperparameter choices are confirmed by getting reasonable results on training data (as illustrated in Figure \ref{fig_birdExample}(b)-(d)) without considering the results on test data (Table \ref{tbl_confMat_avian}).  One direction of future research is automation of hyperparameter setting. During the training phase, we utilize the labeled notes to find clusters which are illustrated in Figure \ref{fig_birdExample}(b). Based on comparing test $\z$-representations to training $\z$-representations, we assign 3 soft scores to $\y_t$ where each score represents $\y_t$'s degree of membership to the corresponding training class (Figure \ref{fig_birdExample}(c)). For this last step, we use a fusion of OCGPs (Section \ref{sec_post}). To make crisp decisions from soft scores of OCGPs, threshold $\tau = 6$ is used.  All predictions below $\tau$ were discarded.  Then all predictions that lasted for less than $20$ time steps were discarded since a note cannot last that short. Note that these last hyperparameters (number of time steps $=20$ and $\tau=6$) are hyperparameters of the system, not that of HLDS.  This is a simple example of how expert knowledge can be used to make crisp decisions from soft classification scores.  An example of final class assignments is illustrated in Figure \ref{fig_birdExample}(d).
The resulting test confusion matrix is illustrated in Table \ref{tbl_confMat_avian}. We wish all instances of the training classes be assigned their true labels at the test time and the rest of the notes to be detected as outliers.  {\tt babico-D} notes are often mispredicted in bioacoustics softwares because of their broad band even in noiseless recordings. The perfect prediction by our system is encouraging.  Note that {\tt babico-F} is not confused with {\tt babico-D}. Bioacoustics softwares have trouble distinguishing them.  Species {\tt babico} and {\tt pocaro} should be indistinguishable wrt their {\tt Z} notes because across species this important signal (of imminent attack by a hawk) is highly canalized.  Therefore, the confusion between {\tt babico-Z} and {\tt pocaro-Z}, observed in our results, is desired. The misprediction of {\tt pocaro-D} notes can potentially be prevented at different stages of our system.  Examples are increasing the dimension of the $\z$-layer to increase separability between mappings from different notes and/or using expert knowledge in post-processing the results of OCGP. We decided to leave such application-dependent detail out in this manuscript.  Another issue that can be resolved at post-processing, is that the beginning of all notes are found with delay.
\section{Experimenting with trumpet notes}
 \label{sec_exp_trumpet}\label{data_trumpet}
 There are $35$ trumpet notes taken from the University of Iowa Musical Instrument Samples (MIS):
\begin{table}
 \begin{tabular}{|c|cccc|}
 \hline
               &{ babico-D}&{ babico-F}&{ pocaro-Z}&outlier\\
            \hline    
{ babico-D}&13/13&0&0&0\\
{ babico-F}&0&4/4&0&0\\
{ pocaro-Z}&0 &0 &2/2 &0\\
\hline
{ babico-A} & 0 &0 &0 &4/4\\
{ babico-Z}& 0 & 0 & 3/4 & 1/4\\
{ pocaro-C}& 0 &0 &0 &6/6\\
{ pocaro-D}&1/7 & 4/7 & 0 & 2/7\\
{ pocaro-E}& 0 &0 &0 &10/10\\
\hline
 \end{tabular}
 \caption{Test confusion matrix for experimenting with avian notes.  The row titles are the true labels and the column titles are the predictions.}
 \label{tbl_confMat_avian}
 \end{table}
 \begin{table}
 \begin{tabular}{|c|cccc|}
 \hline
               &{\tt E5}&{\tt F5}&{\tt G5}&outlier\\
            \hline    
{\tt E5}&4/4&0&0&0\\
{\tt F5}&0&4/4&0&0\\
{\tt G5}&0 &0 &4/4 &0\\
\hline
{outlier} & 0 &0 &0 &32/32\\
\hline
 \end{tabular}
 \caption{Test confusion matrix for experimenting with trumpet notes from MIS dataset.   Outliers are all 32 notes other than {\tt E5}, {\tt F5} and {\tt G5}.}
 \label{tbl_confMat_trumpet}
 \end{table}
 {\tt A3-A5}, {\tt B3-B5}, {\tt C4-C6}, {\tt D4-D6}, {\tt E3-E5}, {\tt F3-F5}, {\tt G3-G5}, {\tt A3s-A5s}, {\tt C4s-C6s}, {\tt D4s-D5s}, {\tt F3s-F5s} and {\tt G3s-G5s}.
 The training clip consists of one instance of each of the {\tt E5}, {\tt F5} and {\tt G5} notes with added noise.  
The test clip consists of all the trumpet notes in a random order.  The system should be tested with different instances of the training notes, but the dataset consists of only one instance of each note.  Therefore, for each {\tt E5}, {\tt F5} and {\tt G5} notes, we created another $4$ instances by adding noise to the original nearly noiseless instances from the dataset. We repeat the experiment for $\sigma_\mathrm{noise} = 0,\frac{1}{10}, \frac{1}{9}, \dots, \frac{1}{3}$ where $\sigma_\mathrm{noise}$ denotes the standard deviations of the added noise.
The time domain sliding windows are preprocessed in the same manner as with the avian dataset. Also all the hyperparameters and post-processing are set as with the avian dataset.  The resulting test confusion matrix with no added noise is illustrated in Table \ref{tbl_confMat_trumpet}.  This perfect result on trumpet notes is due to their stationarity.  For $\sigma_\mathrm{noise}= \frac{1}{10}$ to $\frac{1}{6}$, note {\tt C6s} gets mispredicted as {\tt E5}.  For $\sigma_\mathrm{noise}=\frac{1}{5}$ and $\frac{1}{4}$, {\tt C6} note gets mispredicted as {\tt G5}.  At $\sigma_\mathrm{noise}=\frac{1}{3}$, the distinguishability between the 3 notes is mostly lost.
\section{Conclusion}
We described a system for simultaneous segmentation and classification of notes from recorded audio, in which HLDS was used to model the time-series representing notes.  The results of finding avian notes from recorded audio is promising. We proposed a novel method for parameter setting of HLDS.  In future, we will compare HLDS with constrained non-negative matrix factorization which achieves spectral decomposition for multiple pitch estimation with state-space larger than the output space \cite{vincent2009adaptive}. 
 
\section{Future research}
\label{sec_last}
Here is an outline of possible directions in which future research can be conducted:
\begin{itemize}
\item Incorporating nonlinearity to HLDS in a principled fashion. Some efforts have been conducted by incorporating correntropy \cite{liu2007correntropy} to HLDS \cite{singh2018correntropy}.
\item Automation of hyperparameter setting
\item Learning innovation and error parameters according to data (Section \ref{sec_noiseParams}) and
\item Learning coupling matrices, instead of setting them to fixed values,  but with more constraints to avoid overfitting
\end{itemize}

\vfill\pagebreak

\newpage
\bibliographystyle{ieeetr}
\bibliography{2022_arxiv_HLDS}

\begin{thebibliography}{1}

\bibitem{cinar2018hierarchical}
G.~T. Cinar, P.~M. Sequeira, and J.~C. Principe, ``Hierarchical linear
  dynamical systems for unsupervised musical note recognition,'' {\em Journal
  of the Franklin Institute}, vol.~355, no.~4, pp.~1638--1662, 2018.

\bibitem{coviello2010time}
E.~Coviello, A.~B. Chan, and G.~Lanckriet, ``Time series models for semantic
  music annotation,'' {\em IEEE Transactions on Audio, Speech, and Language
  Processing}, vol.~19, no.~5, pp.~1343--1359, 2010.

\bibitem{chan2009layered}
A.~B. Chan and N.~Vasconcelos, ``Layered dynamic textures,'' {\em IEEE
  Transactions on Pattern Analysis and Machine Intelligence}, vol.~31, no.~10,
  pp.~1862--1879, 2009.

\bibitem{PCKMD}
L.~Kalantari, J.~Principe, and K.~Sieving, ``Kernel mahalanobis distance with
  automated parameter setting for multiclass classification.'' {\it Submitted
  to IJCNN 2020}.

\bibitem{vincent2009adaptive}
E.~Vincent, N.~Bertin, and R.~Badeau, ``Adaptive harmonic spectral
  decomposition for multiple pitch estimation,'' {\em IEEE Transactions on
  Audio, Speech, and Language Processing}, vol.~18, no.~3, pp.~528--537, 2009.

\bibitem{liu2007correntropy}
W.~Liu, P.~P. Pokharel, and J.~C. Pr{\'\i}ncipe, ``Correntropy: Properties and
  applications in non-gaussian signal processing,'' {\em IEEE Transactions on
  Signal Processing}, vol.~55, no.~11, pp.~5286--5298, 2007.

\bibitem{singh2018correntropy}
R.~Singh and J.~C. Principe, ``Correntropy based hierarchical linear dynamical
  system for speech recognition,'' in {\em 2018 International Joint Conference
  on Neural Networks (IJCNN)}, pp.~1--7, IEEE, 2018.

\end{thebibliography}

\end{document}